\documentclass[reprint,amsmath,amssymb]{revtex4-1}
\usepackage{graphicx,grffile}
\usepackage{dcolumn}
\usepackage{bm}
\usepackage{subfig}
\usepackage{caption}
\usepackage{xcolor}
\usepackage{float}

\begin{document}

\title{Suppression and Revival of Oscillations through Time-varying Interaction}
\author{Sudhanshu Shekhar Chaurasia$^{1}$}

\author{Anshul Choudhary$^{1,2}$}
 
\author{Manish Dev Shrimali$^3$}
 
\author{Sudeshna Sinha$^1$}
\affiliation{
                \vspace{1 mm}$^1$Indian Institute of Science Education and Research (IISER) Mohali,\\
                Knowledge City, SAS Nagar, Sector 81, Manauli PO 140 306, Punjab, India
                }%

\affiliation{
                \vspace{1 mm}$^2$Institute for Chemistry and Biology of the Marine Environment (ICBM), University of Oldenburg, Carl-von-Ossietzky-Strasse 9, 26129 Oldenburg, Germany
                }%

\affiliation{
                $^3$Department of Physics, Central University of Rajasthan,\\
                 Kishangarh, Ajmer 305 817, India
                }%

\begin{abstract}
  We explore the dynamical consequences of switching the coupling form
  in a system of coupled oscillators. We consider two types of
  switching, one where the coupling function changes periodically and
  one where it changes probabilistically. We find, through bifurcation
  diagrams and Basin Stability analysis, that there exists a window in
  coupling strength where the oscillations get suppressed.  Beyond
  this window, the oscillations are revived again. A similar trend
  emerges with respect to the relative predominance of the coupling
  forms, with the largest window of fixed point dynamics arising where
  there is balance in the probability of occurrence of the coupling
  forms. Further, significantly, more rapid switching of coupling
  forms yields large regions of oscillation suppression. Lastly, we
  propose an effective model for the dynamics arising from switched
  coupling forms and demonstrate how this model captures the basic
  features observed in numerical simulations and also offers an
  accurate estimate of the fixed point region through linear stability
  analysis.
\end{abstract}

\maketitle

\section{Introduction}

Coupled dynamical systems have been extensively studied over the last
few decades as they provide us a framework for modelling many complex
systems \cite{bio_osc,ecosystem_own}. The focus of investigations have mostly been
on phenomena emerging under variation of the local dynamics of the
units and the interactions among them. Most studies have assumed the
interactions among the nodes to be invariant over time. However in
recent times there have been efforts to incorporate a time-varying
links, namely changing connections between the units in a dynamical
network
\cite{links1,links2,links3,links4,links5,links6,links7,links8,links9,links10}. Such
time varying interactions model the evolution of connections over
time, and are commonly found in physical, biological, social and
engineered systems
\cite{timevarying1,timevarying2,timevarying3,timevarying4,timevarying5}.
Studies so far have considered the variation in links as a function of
time, while the form of coupling remains the same. Here we will
explore a new direction in time-varying interactions: we will study
the {\em effect of switched coupling forms} on the emergent
behaviour. 

We consider two coupling functions, one diffusive and the other
conjugate coupling between the two oscillators. Coupling via conjugate
variables is natural in a variety of experimental situations where
sub-systems are coupled by feeding the output of one into the
other. An example from the recent literature is provided by the
experiments of Kim and Roy on coupled semiconductor laser systems
\cite{laser1}, where the photon intensity fluctuation from one laser
is used to modulate the injection current of the other, and vice
versa.  Hybrid coupling \cite{hybrid1,hybrid2,hybrid3} also has
relevance in ecological models, where migration and cross-predation
(analogous to conjugate coupling) \cite{conj_coup} occurs between two
population patches, namely over some time migration or diffusive
coupling may be dominant, while at other times cross-predation between
the two patches is prevalent.

The primary goal of this study is to demonstrate the non-trivial
dynamical states arising out of the temporal interplay between two
coupling forms. The test-bed of our inquiry will be a generic system
of coupled oscillators, which we describe in detail below.

\section{Coupled oscillators}

	A general form of coupled dynamical oscillators is given by: 
	\begin{eqnarray}
		\dot{X}_i = F(X_i) + K G_i(X_j,X_j',X_i)
	\end{eqnarray}	
where $X_i$ denotes the set of $m$ dynamical variables of the $i$th oscillator. The matrix $K$ of dimension $m \times m$ contains information on the coupling topology. $G_i$ is the coupling function that represents the nature of the interaction and the variables involved in the interaction term, with the superscript primes ($'$) on $X$ denoting conjugate or ``dissimilar'' variables \cite{mixed}.

Now, complex systems often undergo Hopf bifurcations and sufficiently close to
such a bifurcation point, the variables which have slower time-scales
can be eliminated. This leaves us with a couple of simple first order
ordinary differential equations, popularly known as the Stuart-Landau
system \cite{kuramoto}. In this work we will consider two diffusively
coupled oscillators of this form, namely two coupled Stuart-Landau
systems.

Specifically, when the coupling between the oscillators is through {\em
  similar} variables, the dynamical equations are given by:
	\begin{eqnarray}
		\label{eqn_similar}
		\dot{x_1}&=& f^x(x_1,y_1) + \varepsilon(x_2-x_1) \nonumber \\
		\dot{y_1}&=& f^y(x_1,y_1) \\
		\dot{x_2}&=& f^x(x_2,y_2) + \varepsilon(x_1-x_2) \nonumber \\
		\dot{y_2}&=& f^y(x_2,y_2) \nonumber
	\end{eqnarray}
and when the oscillators are coupled to each other through
      {\em  dissimilar} variables, the dynamical equations are given by:

	\begin{eqnarray}
		\label{eqn_dissimilar}
		\dot{x_1}&=& f^x(x_1,y_1) + \varepsilon(y_2-x_1) \nonumber \\
		\dot{y_1}&=& f^y(x_1,y_1) \\
		\dot{x_2}&=& f^x(x_2,y_2) + \varepsilon(y_1-x_2) \nonumber \\
		\dot{y_2}&=& f^y(x_2,y_2) \nonumber
	\end{eqnarray}
with

	\begin{eqnarray}
		\label{eqn_stuart}
		f^x&=& x ( 1- (x^2 + y^2 ) ) -5y \\
		f^y&=& y ( 1- (x^2 + y^2 ) ) +5x \nonumber
	\end{eqnarray}

Coupled Stuart-Landau oscillators have been a good model for the study of oscillation death \cite{dissimilar1,dissimilar2}  and amplitude death \cite{ad1,ad2,ad3,ad4}. When the coupling is through similar variables (cf. Eqn.~\ref{eqn_similar}), the oscillators get synchronized at a very low coupling strengths, and remain oscillatory and synchronized up to very high coupling strength. However when coupling is through dissimilar variables (cf. Eqn.~\ref{eqn_dissimilar}), the system shows oscillatory behaviour for small coupling strengths, and increasing coupling strengths give rise to amplitude death and subsequently oscillation death \cite{dissimilar3}.

\section{Time-varying Coupling Form}

Now we consider a scenario where the form of the coupling between the
oscillators is time dependent, namely, the form of coupling {\em
  switches between the similar and dissimilar variables}. The
switching of the coupling form may be periodic or probabilistic.

\subsection{Periodic Switching of Coupling Forms} 

Here the oscillators change their coupling form periodically. If the
time period of the switch is $T$, we consider the system to be coupled
via similar variables for fraction $f_{sim}$ of the cycle, followed by
coupling to dissimilar variables for the remaining part. So when
$f_{sim}=0$, the oscillators are always coupled to dissimilar
variables and when $f_{sim}=1$ the oscillators are coupled through
similar variable for all time. For $0 < f_{sim} < 1$, the oscillators
experience coupling through similar variables for time $f_{sim} T$,
followed by coupling through dissimilar variables for time $T (1 -
f_{sim})$, in each cycle of period $T$.

	

\begin{figure}[H]
		\centering
		\includegraphics[width=1.0\linewidth]{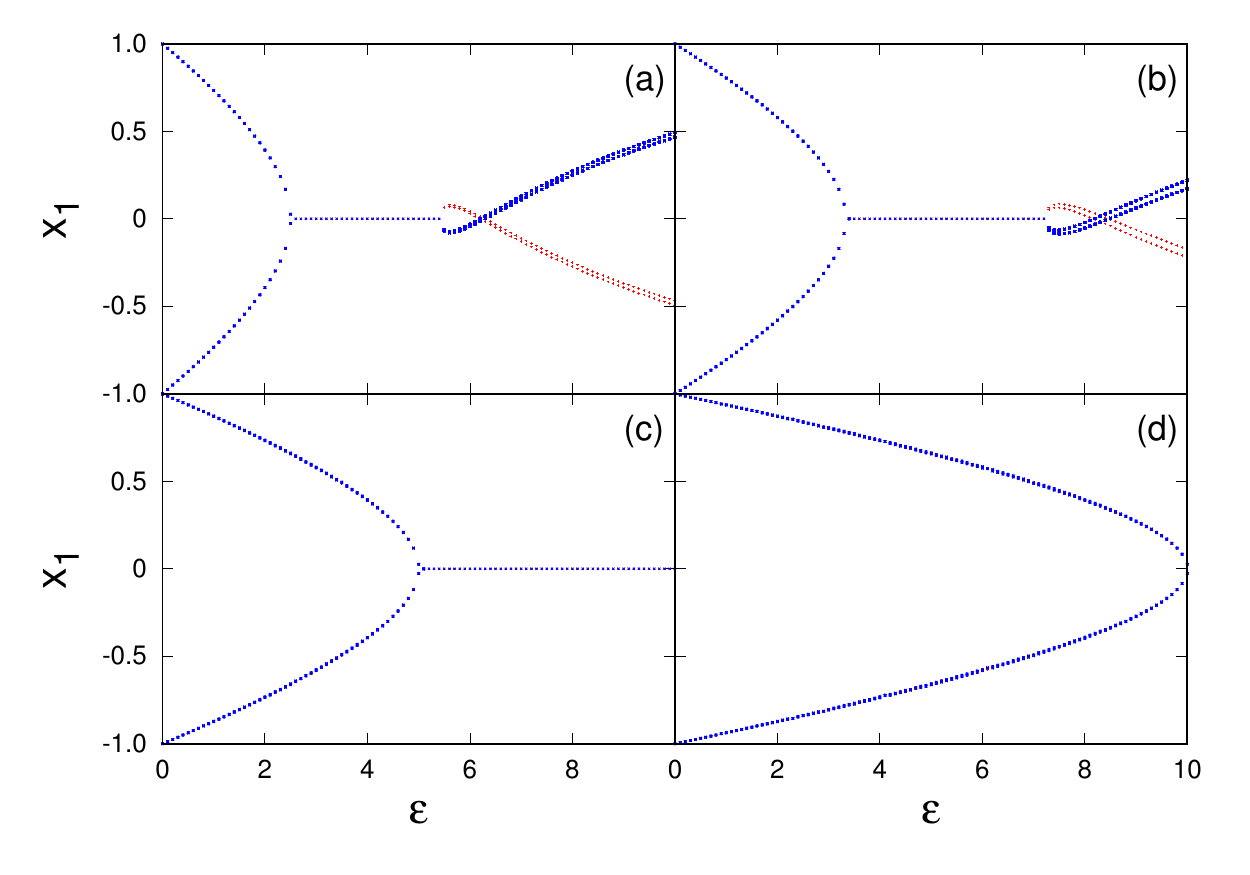}
		\caption{Bifurcation diagram displaying variable $x$
                  of one of the oscillators, with respect to coupling
                  strength $\varepsilon$. Here the coupling switches
                  between similar and dissimilar variables
                  periodically, with similar-variable coupling (for
                  time $f_{sim} T$) followed by dissimilar-variable
                  coupling (for time $(1-f_{sim})T$), for (a)
                  $f_{sim}$=$0.2$, (b) $f_{sim}$=$0.4$, (c)
                  $f_{sim}$=$0.6$ and (d) $f_{sim}$=$0.8$, and time
                  period of switching $T=0.02$. The two colours
                  represent symmetric solutions around the unstable fixed
                  point, arising from different initial conditions.}
		\label{fig:bif_periodic_fixed_T}
	\end{figure}

Fig.~\ref{fig:bif_periodic_fixed_T} shows the bifurcation diagram for
a system of two coupled oscillators with periodically changing
coupling form, with respect to coupling strength, for different
$f_{sim}$. At $\varepsilon=0$, i.e. for uncoupled Stuart-Landau
oscillators, one naturally obtains period $1$ oscillations. Increasing
the coupling strength results in suppression of oscillations.
Interestingly though, the window of coupling strength over which
oscillations are suppressed depends non-monotonically on $f_{sim}$. At
first, as $f_{sim}$ increases the fixed point window increases
(cf. Fig.~\ref{fig:bif_periodic_fixed_T}a for $f_{sim}=0.2$ vis-a-vis
Fig.~\ref{fig:bif_periodic_fixed_T}c at $f_{sim}=0.6$). However, when
$f_{sim}$ gets even larger this window vanishes entirely
(cf. Fig.~\ref{fig:bif_periodic_fixed_T}d), namely the oscillations
are no longer suppressed anywhere.

\begin{figure}[H]
		\centering
		\includegraphics[width=0.85\linewidth]{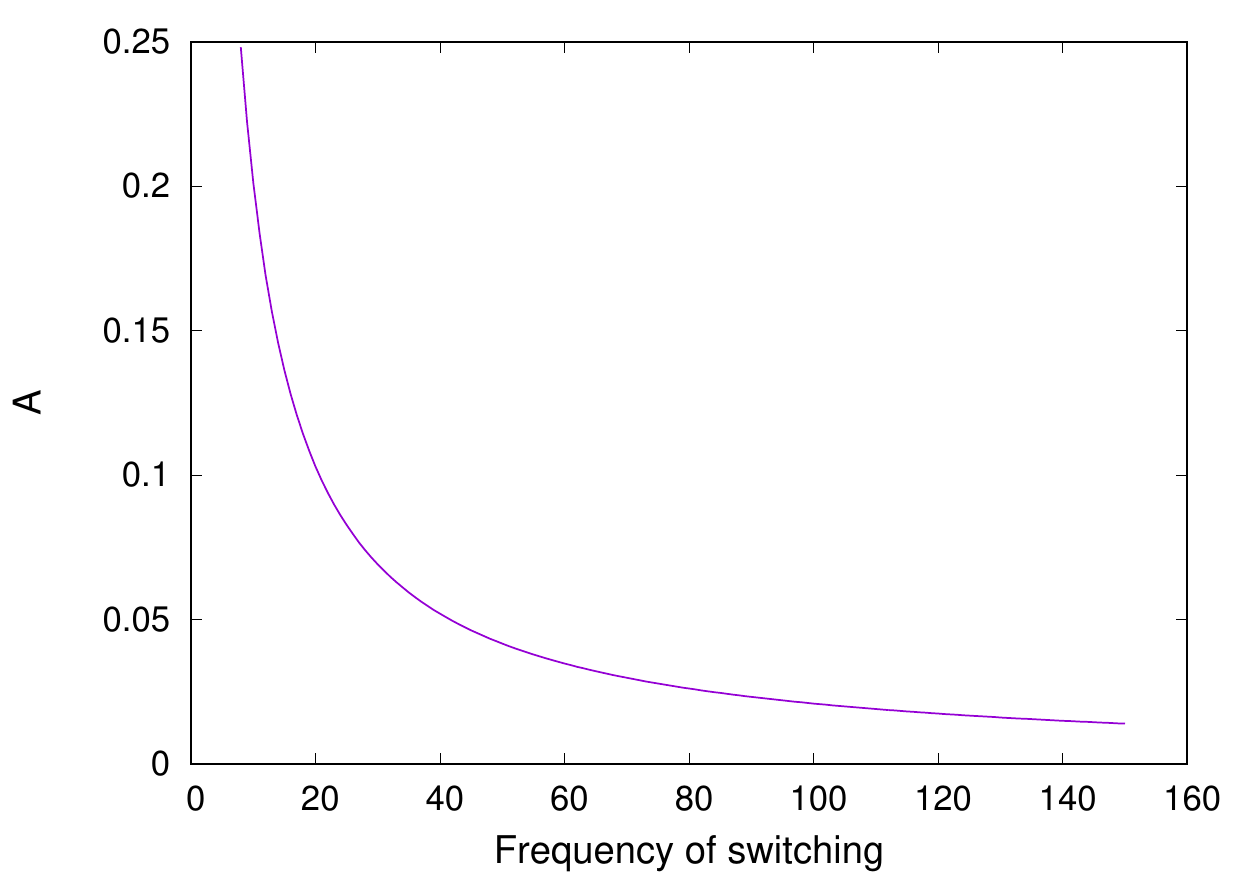}
		\caption{Dependence of $A = x_{max}-x_{min}$  on the frequency of switching (where $x_{max}$/$x_{min}$ are the maximum/minimum values of variable $x$ of an oscillator). Here $f_{sim}=0.4$, $\varepsilon=9.0$, and the coupling switches between similar and dissimilar variables periodically, with similar-variable coupling for time $f_{sim} T$, followed by dissimilar-variable coupling for time $(1-f_{sim})T$, where $1/T$ is the frequency of switching. }
		\label{fig:osc_amplitude_periodic}
	\end{figure}

	Further, notice from Fig.~\ref{fig:bif_periodic_fixed_T}a-b that low-amplitude oscillations are restored at higher coupling strengths again, for intermediate $f_{sim}$. Fig.~\ref{fig:osc_amplitude_periodic} shows the effect of the frequency of switching coupling forms upon the amplitude of the revived oscillations. Increase in frequency of switching leads to the reduction of oscillation amplitude, and the results approach those arising from effective mean-field like dynamical equations (cf. Eqn. \ref{eqn_effective}) which will be presented in Section V (namely, Fig.~\ref{fig:bif_periodic_fixed_T}(b) approaches Fig.~\ref{fig:bif_effective_fixed_fsim}(b) obtained from an approximate effective description of the system).
	
\subsection{Probabilistic Switching of Coupling Forms}
		
Here the oscillators change their coupling form at intervals of time
$T$, with the coupling form chosen probabilistically. We consider the
probability for the oscillators to be coupled via similar variables to
be $p_{sim}$, and the probability of coupling mediated via dissimilar
variables to be $(1-p_{sim})$. For $0 < p_{sim} < 1$, at the time of
switching, the similar-variable coupling form is chosen with
probability $p_{sim}$ and the dissimilar-variable form is chosen with
probability $(1-p_{sim})$. So larger $p_{sim}$ favours coupling
through similar variables and smaller $p_{sim}$ favours
dissimilar-variable coupling, with the oscillators always experiencing
dissimilar-variable coupling for the limiting case of $p_{sim}=0$ and
similar-variable coupling for $p_{sim}=1$. Here the probability
$p_{sim}$ plays a role equivalent to $f_{sim}$ in the case of periodic
switching of coupling forms.

\begin{figure}[H]
			\centering
			\includegraphics[width=1.0\linewidth]{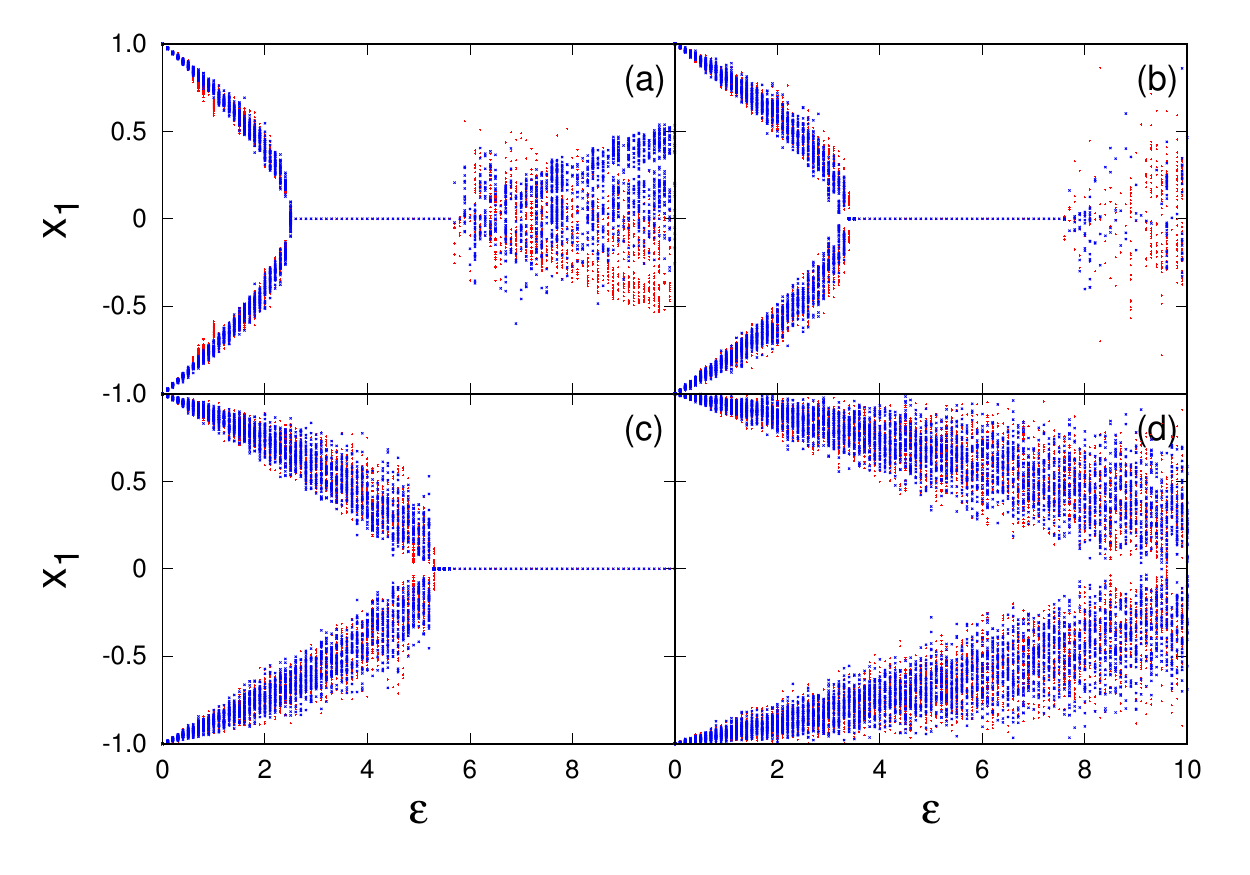}
			\caption{Bifurcation diagram displaying
                          variable $x$ of one of the oscillators, with
                          respect to coupling strength
                          $\varepsilon$. Here the coupling form
                          probabilistically switches at time intervals
                          of $T$ between similar and dissimilar
                          variables, with the probability of
                          similar-variable coupling being $p_{sim}$,
                          and the probability of dissimilar-variable
                          coupling being $(1-p_{sim})$, for (a)
                          $p_{sim}=0.2$, (b) $p_{sim}=0.4$, (c)
                          $p_{sim}=0.6$ and (d) $p_{sim}=0.8$ (with
                          $T=0.02$). The two colours represent
                          emergent dynamics from different initial
                          conditions}
			\label{fig:bif_probabilistic_fixed_T}
		\end{figure}

\begin{figure}[H]
			\centering
			\includegraphics[width=0.85\linewidth]{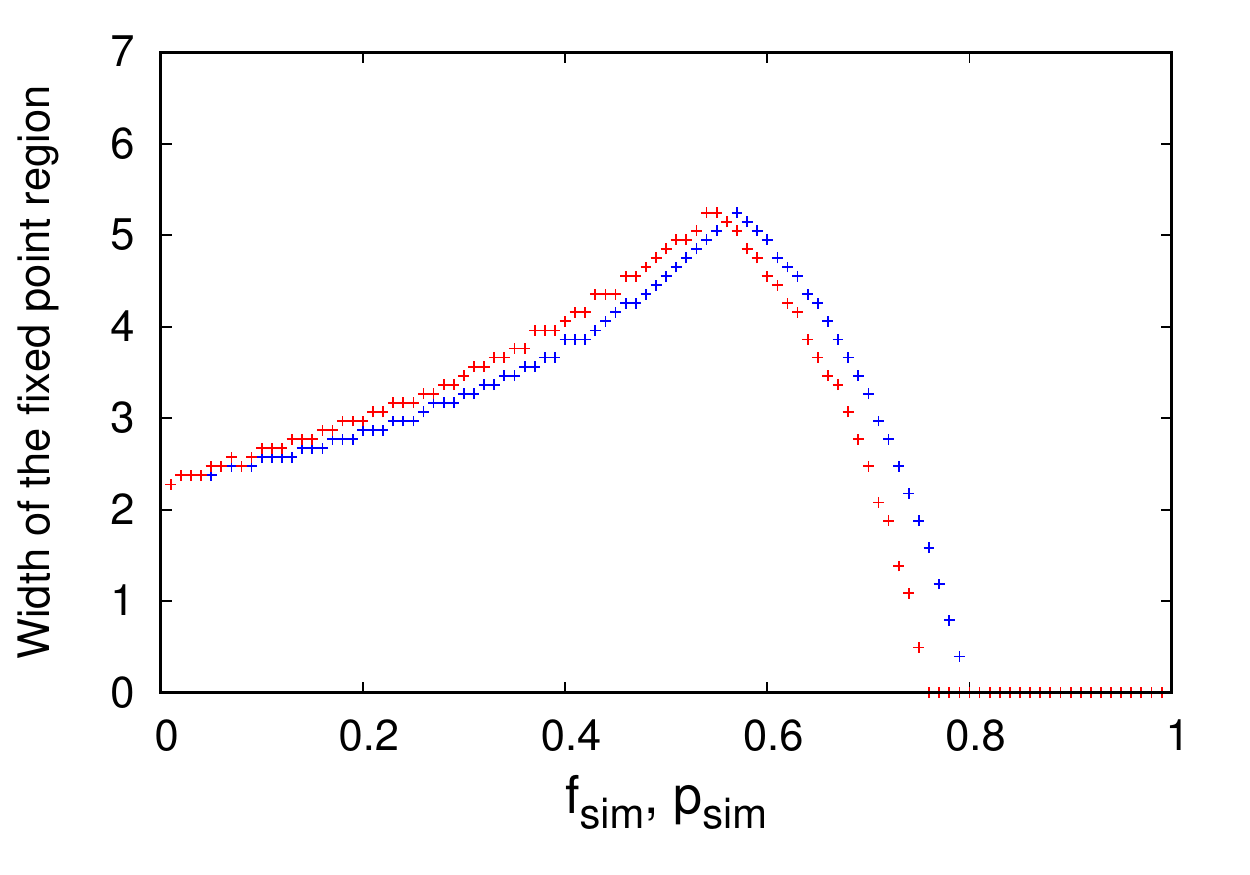}
			\caption{Width of the fixed point window as a function of $f_{sim}$ (blue) for the case of periodic switching of coupling forms, and $p_{sim}$ (red) for the case of probabilistic switching of coupling forms, with switching period $T=0.02$.}
			\label{fig:fp_width_t_switch_0.02} 
		\end{figure}

Fig.~\ref{fig:bif_probabilistic_fixed_T} shows the bifurcation diagram,
with respect to coupling strength, for different $p_{sim}$. Again it
is evident that the window of coupling strength over which
oscillations are suppressed, depends non-monotonically on $p_{sim}$,
as it did under variation of $f_{sim}$ for the case of periodically
switched coupling forms
(cf. Fig.~\ref{fig:fp_width_t_switch_0.02}). First, as $p_{sim}$
increases from zero, the fixed point window increases, as seen from
Fig.~\ref{fig:bif_probabilistic_fixed_T}a for $p_{sim}=0.2$ vis-a-vis
Fig.~\ref{fig:bif_probabilistic_fixed_T}c for $p_{sim}=0.6$. However,
when $p_{sim}$ gets even larger this window vanishes entirely, as
evident from Fig.~\ref{fig:bif_probabilistic_fixed_T}d, and the
oscillations are no longer suppressed anywhere. This suggests that
when the probability of coupling through similar variables and
dissimilar variables is similar (i.e. $p_{sim} \sim 0.5$) oscillations
are suppressed to the greatest degree.
		
		Fig.~\ref{fig:fp_width_t_switch_0.02} shows the width
                of fixed point window, with respect to $f_{sim}$ for
                the case of periodic switching of coupling forms, and
                $p_{sim}$ for the case of probabilistic switching of
                coupling forms. As shown, the width of the fixed point
                window is non-monotonic, and has a maximum at $f_{sim}
                (p_{sim}) \sim0.58$. Namely, the largest window of
                fixed point dynamics arises where there is balance in
                the probability of occurrence of the coupling forms.
		
\section{Global stability measure}

The commonly employed linear stability analysis, based on the
linearization in the neighbourhood of fixed points, provides only local
information about the stability at the fixed point. It cannot
accurately indicate the stability for large perturbations, nor the
basin of attraction of the dynamics, especially in the presence of
other attractors in phase space.

Here we calculate the Basin Stability of the dynamical states
\cite{basin}. This is a more robust and global estimate of stability,
and effectively incorporates non-local and non-linear effects on the
stability of fixed points. Specifically, Basin Stability is calculated
as follows: we choose a large number of random initial conditions,
spread uniformly over a volume of phase space, and find what fraction
of these are attracted to stable fixed points.

\begin{figure}[H]
			\centering
			\includegraphics[width=1.0\linewidth]{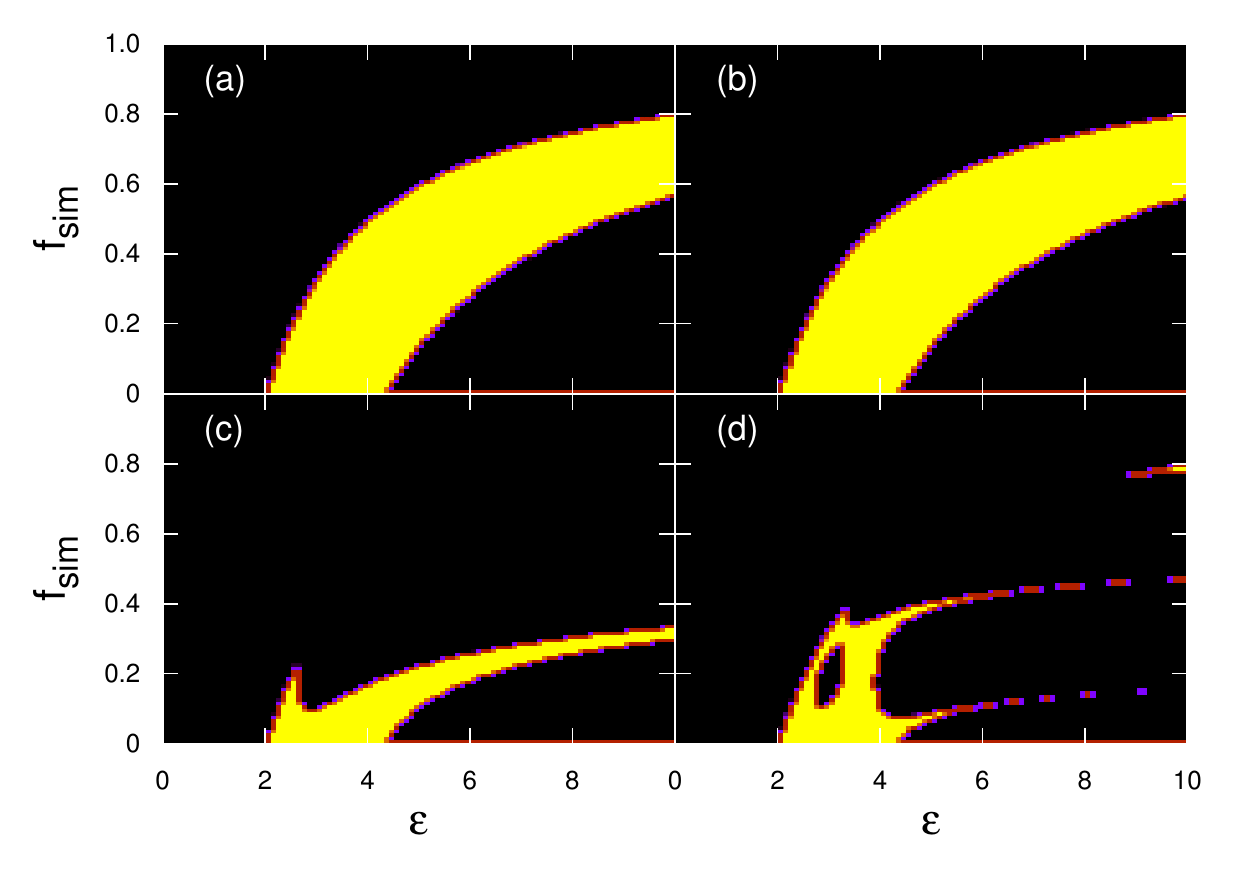}
			\caption{Basin Stability of the fixed point state of coupled oscillators, in the parameter space of coupling strength $\varepsilon$ and $f_{sim}$. Here the coupling periodically switches between similar and dissimilar variables, with similar-variable coupling (for time $f_{sim} T$), followed by dissimilar-variable coupling (for time $(1-f_{sim})T$), for (a) $T=0.01$, (b) $T=0.10$, (c) $T=1.00$ and (d) $T=2.00$.  The region in yellow represents fixed point dynamics (i.e. oscillation suppression), and the region in black represents oscillatory dynamics.}
			\label{fig:bs_periodic_fixed_t_switch}
		\end{figure}

Figs~\ref{fig:bs_periodic_fixed_t_switch} shows the Basin Stability of
the fixed point state, in the parameter space of $f_{sim}$ and
coupling strengths, for different time periods of switching
$T$. Clearly one obtains oscillation suppression in windows of
coupling strength and $f_{sim}$, and oscillation revival again beyond
the window. The window of coupling strengths that gives rise to fixed
points is very sensitive to the frequency of switching, at low
frequencies.  After a high enough switching frequency (i.e. low enough
$T$), the fixed point region remains unchanged, as evident through the
fact that Fig.~\ref{fig:bs_periodic_fixed_t_switch}a and
Fig.~\ref{fig:bs_periodic_fixed_t_switch}b are identical.

Now, the dependence of the fixed point window on $f_{sim}$ is actually
quite counter-intuitive, as already indicated in the bifurcation
diagrams. For rapidly switched coupling forms, at large coupling
strengths, the oscillation suppression occurs at an {\em intermediate}
value of $f_{sim}$. Namely, as the dominance of similar-variable
coupling $f_{sim}$ increases the oscillations are first suppressed and
then after a point the oscillations are revived again, with the window
of fixed points shifting towards higher $f_{sim}$, as coupling
strength increases.  This is counter-intuitive, as similar-variable
coupling is known to only allow oscillations, while
dissimilar-variable coupling can yield some windows of oscillation
suppression. Also interestingly for low $f_{sim}$, as we increase the
coupling strength, first we encounter oscillation suppression and then
on further increase of coupling strength the oscillations are
restored. So there exists an intermediate window of coupling strength
that yields fixed point dynamics.

\begin{figure}[H]
			\centering
			\includegraphics[width=1.0\linewidth]{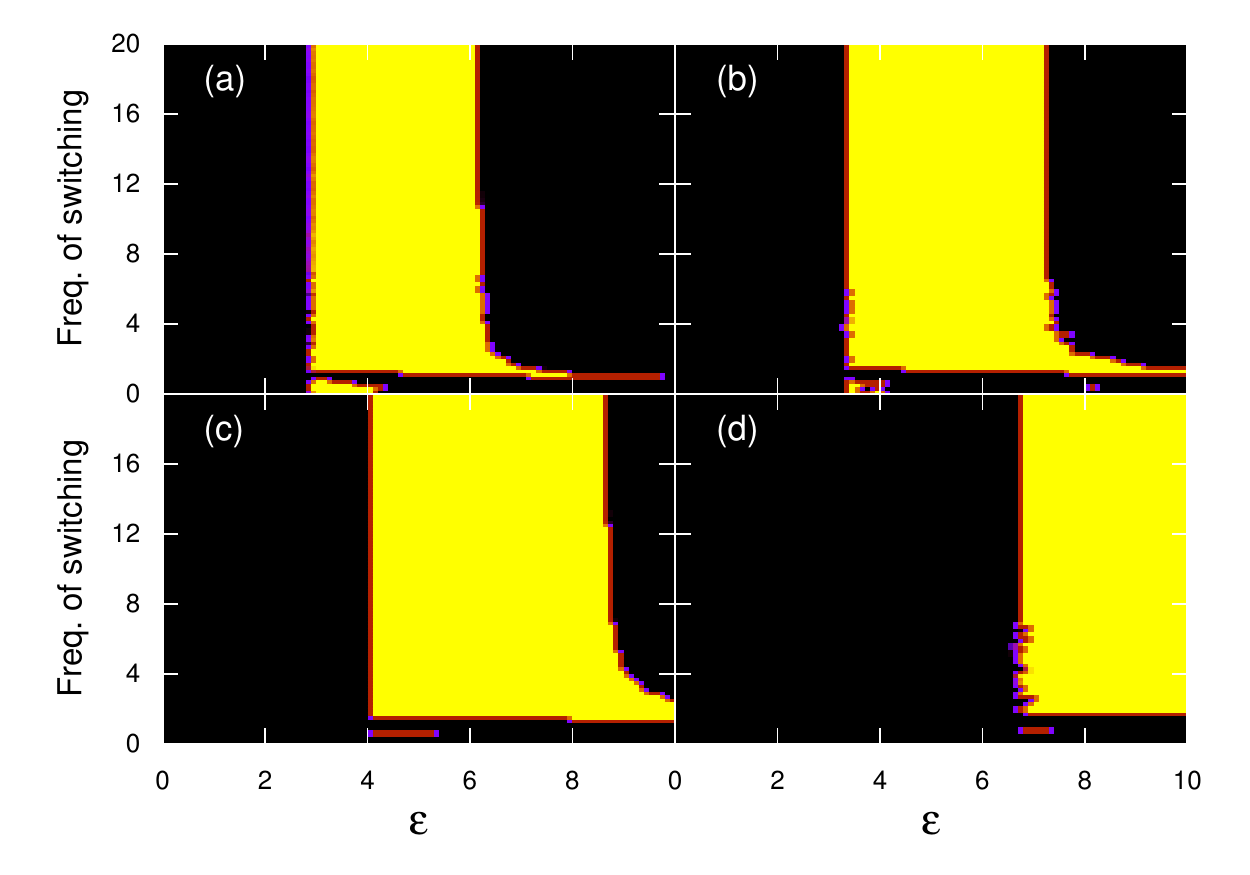}
			\caption{Basin Stability of the fixed point
                          state of coupled oscillators, in the
                          parameter space of coupling strength
                          $\varepsilon$ and frequency of
                          switching. Here the coupling periodically
                          switches between similar and dissimilar
                          variables, with similar-variable coupling
                          (for time $f_{sim} T$), followed by
                          dissimilar-variable coupling (for time
                          $(1-f_{sim})T$), for (a) $f_{sim}=0.3$, (b)
                          $f_{sim}=0.4$, (c) $f_{sim}=0.5$ and (d) $f_{sim}=0.7$.  The
                          region in yellow represents fixed point
                          dynamics (i.e. oscillation suppression), and
                          the region in black represents oscillatory
                          dynamics.}
			\label{fig:bs_periodic_fixed_f_sim}
		\end{figure}

Fig~\ref{fig:bs_periodic_fixed_f_sim} shows the Basin Stability of the
fixed point state, in the parameter space of the frequency of
switching and coupling strengths, for different $f_{sim}$. It is clear
again that after a critical switching frequency the dynamics does not
depend on the rate at which the coupling form is
changed. Significantly, it is also evident that {\em fast changes in
  coupling form, namely lower time period for change, yields large
  fixed point regions in parameter space.} 
However, at low switching frequencies the emergent dynamics is 
sensitive to how rapidly the coupling form varies.

\begin{figure}[H]
			\centering
			\includegraphics[width=1.0\linewidth]{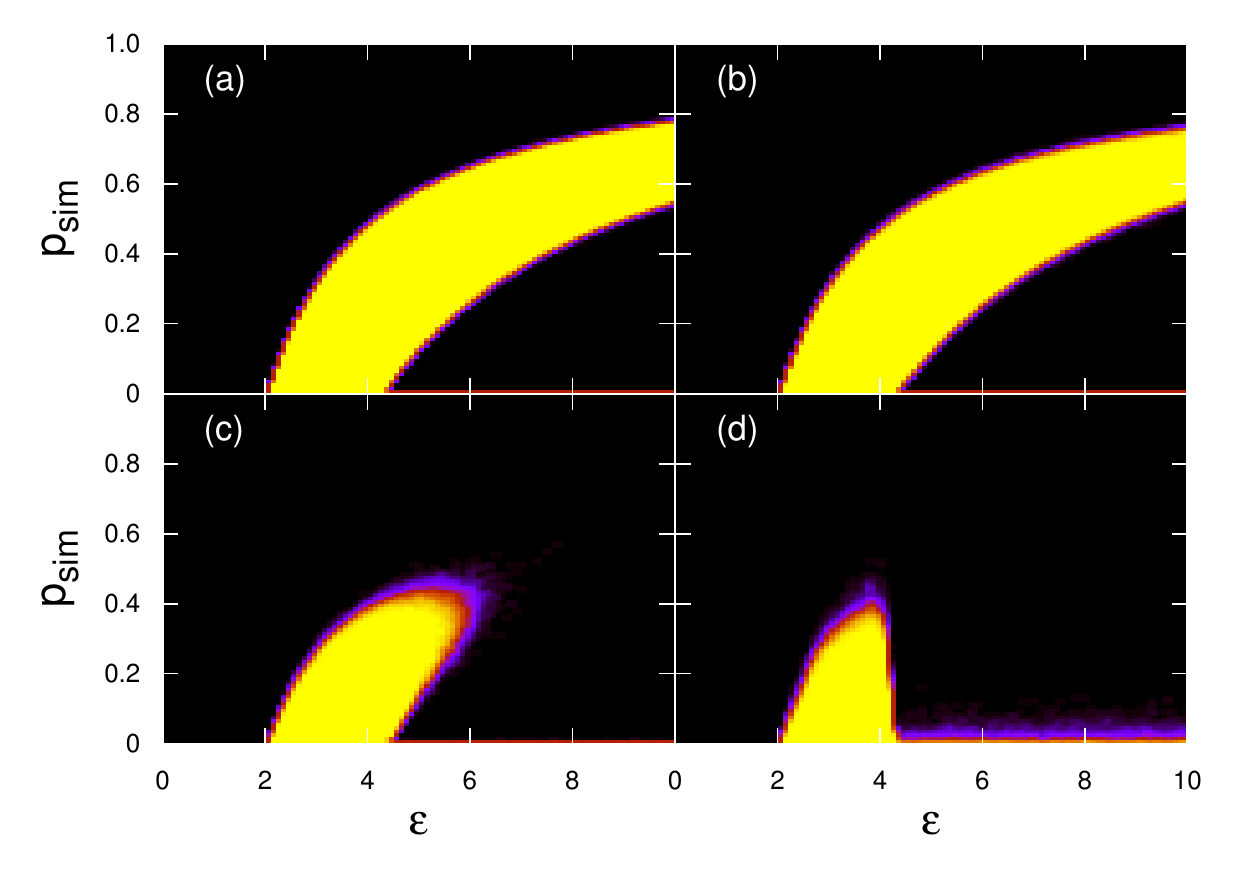}
			\caption{Basin Stability of the fixed point state of coupled oscillators, in the parameter space of coupling strength $\varepsilon$ and $p_{sim}$. Here the coupling probabilistically switches between similar and dissimilar variables, with the probability of similar-variable coupling being $p_{sim}$, and the probability of dissimilar-variable coupling being $(1-p_{sim})$ with the period of switching (a) $T=0.01$, (b) $T=0.02$, (c) $T=0.10$ and (d) $T=1.00$.   The region in yellow represents fixed point dynamics (i.e. oscillation suppression), and the region in black represents oscillatory dynamics. Notice the marked similarity of panel (a) with Fig.~\ref{fig:bs_periodic_fixed_t_switch}a.}
			\label{fig:bs_probabilistic_fixed_t_switch}
		\end{figure}

                Fig~\ref{fig:bs_probabilistic_fixed_t_switch} shows
                the Basin Stability of the fixed point state for the
                case of probabilistically varying coupling form, in
                the parameter space of $p_{sim}$ and coupling
                strengths. {\em Interestingly again, as we increase
                  the coupling strength, oscillations first get
                  suppressed and then restored.} Also notice the
                marked similarity of
                Fig.~\ref{fig:bs_probabilistic_fixed_t_switch}a and
                Fig.~\ref{fig:bs_periodic_fixed_t_switch}a. Namely
                frequent periodic switching of coupling forms yields
                the same result as the frequent probabilistic
                switching of coupling forms.

\begin{figure}[H]
			\centering
			\includegraphics[width=1.0\linewidth]{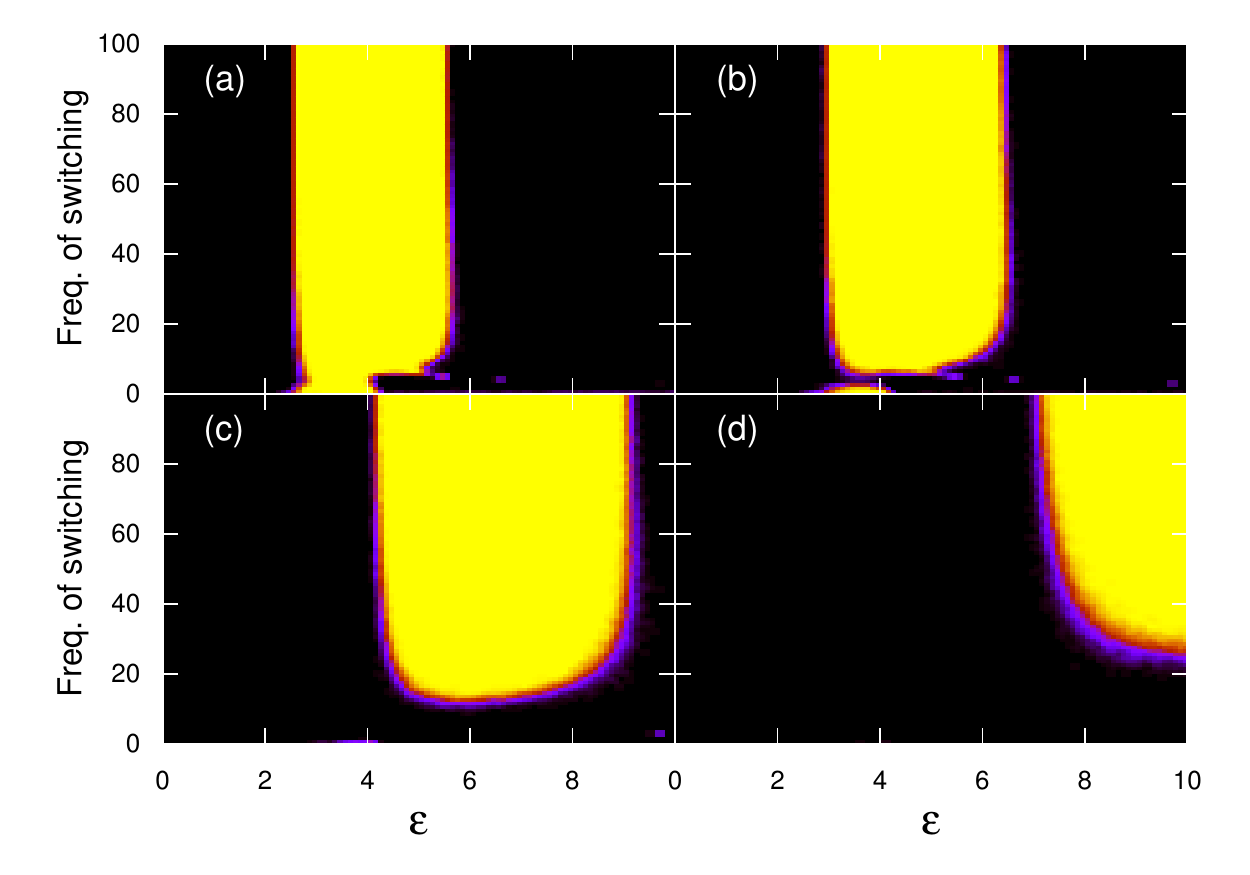}
			\caption{Basin Stability results for coupled oscillators, in the parameter space of coupling strength $\varepsilon$ and frequency of switching, when coupling switches between similar and dissimilar variables probabilistically, with a probability of similar coupling $p_{sim}$, and probability of dissimilar coupling $(1-p_{sim})$, for (a) $p_{sim}=0.2$, (b) $p_{sim}=0.3$, (c) $p_{sim}=0.5$ and (d) $p_{sim}=0.7$. The region in yellow represents fixed point dynamics (i.e. oscillation suppression), and the region in black represents oscillatory dynamics.}
			\label{fig:bs_probabilistic_fixed_p}
		\end{figure}

		Fig.~\ref{fig:bs_probabilistic_fixed_p} shows the
                Basin Stability of the fixed point state, in the
                parameter space of frequency of switching and coupling
                strengths. Clearly the effects of the frequency of
                switching are pronounced over a larger range of
                switching frequency for probabilistic switching, as
                compared to periodic switching. But significantly
                again, it is evident that {\em fast changes in
                  coupling form, namely lower time period for change,
                  yields large fixed point regions in parameter
                  space.} Lastly, it is also clear that as the
                frequency of switching increases, the fixed point
                region moves towards higher values of $p_{sim}$ where
                similar-variable coupling dominates. This is again
                surprising, as similar-variable coupling is known to
                support only oscillations, while dissimilar-variable
                coupling has more propensity towards oscillation
                suppression.

\section{Effective model for time dependent coupling}
	
Now we attempt to rationalize our results through an effective
phenomenological model for the dynamics. The idea is to mimic the time-dependent
coupling by a coupling form where the similar and dissimilar coupling
forms are appropriately weighted by $f_{sim}$. This is given by:
	
	\begin{eqnarray}
		\label{eqn_effective}
		\dot{x_1}&=& f^x(x_1,y_1) + \varepsilon [  f_{sim} (x_2-x_1) + (1-f_{sim})(y_2-x_1) ] \nonumber \\
		\dot{y_1}&=& f^y(x_1,y_1) \\
		\dot{x_2}&=& f^x(x_2,y_2) + \varepsilon [  f_{sim} (x_1-x_2) + (1-f_{sim})(y_1-x_2) ] \nonumber \\
		\dot{y_2}&=& f^y(x_2,y_2) \nonumber
	\end{eqnarray}
	
This effective picture is expected to hold true when the frequency of switching is very high (namely $T$ is very small).  Completely equivalent results can be obtained with $p_{sim}$ in place of $f_{sim}$ in the equations above. 

	\begin{figure}[H]
		\centering
		\includegraphics[width=1.0\linewidth]{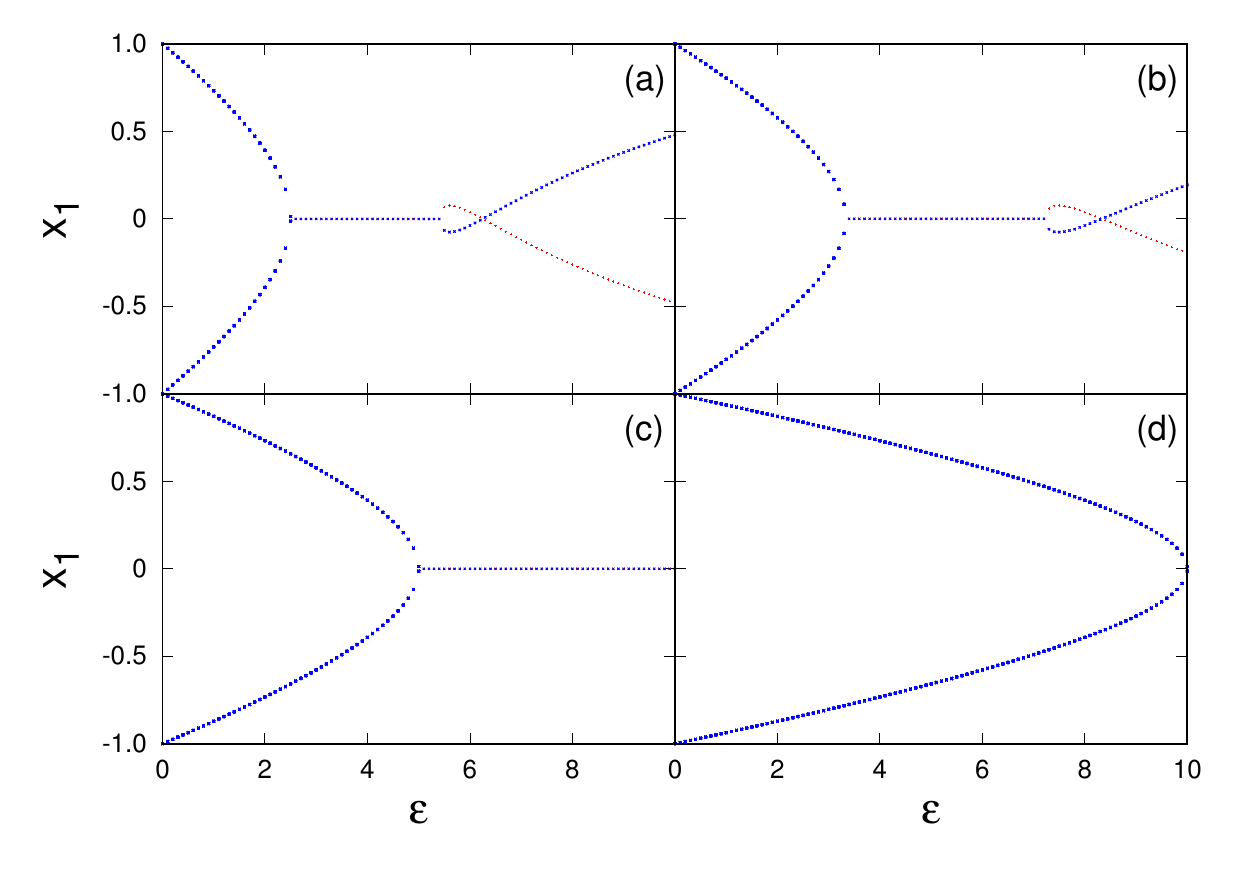}
		\caption{Bifurcation diagram displaying variable $x$ of one of the oscillators of the effective coupled system given by Eqn.~\ref{eqn_effective}, with respect to coupling strength $\varepsilon$, at different $f_{sim}$ values: (a) $f_{sim}$=$0.2$, (b) $f_{sim}$=$0.4$, (c) $f_{sim}$=$0.6$ and (d) $f_{sim}$=$0.8$. The two colours represent symmetric solutions around the unstable fixed point, arising from different initial conditions.}
		\label{fig:bif_effective_fixed_fsim}
	\end{figure}

        Fig.~\ref{fig:bif_effective_fixed_fsim} displays the
        bifurcation diagram of the effective coupled system given by
        Eqn.~\ref{eqn_effective}, and provides insight into the
        suppression and revival of oscillations in the coupled
        oscillator system with time-varying coupling forms. In
        particular, notice the marked similarity of the fixed point
        region in Fig.~\ref{fig:bif_effective_fixed_fsim} with the
        fixed point regions evident in
        Figs.~\ref{fig:bif_periodic_fixed_T} and
        \ref{fig:bif_probabilistic_fixed_T}. 

	\begin{figure}[H]
		\centering
		\includegraphics[width=0.85\linewidth]{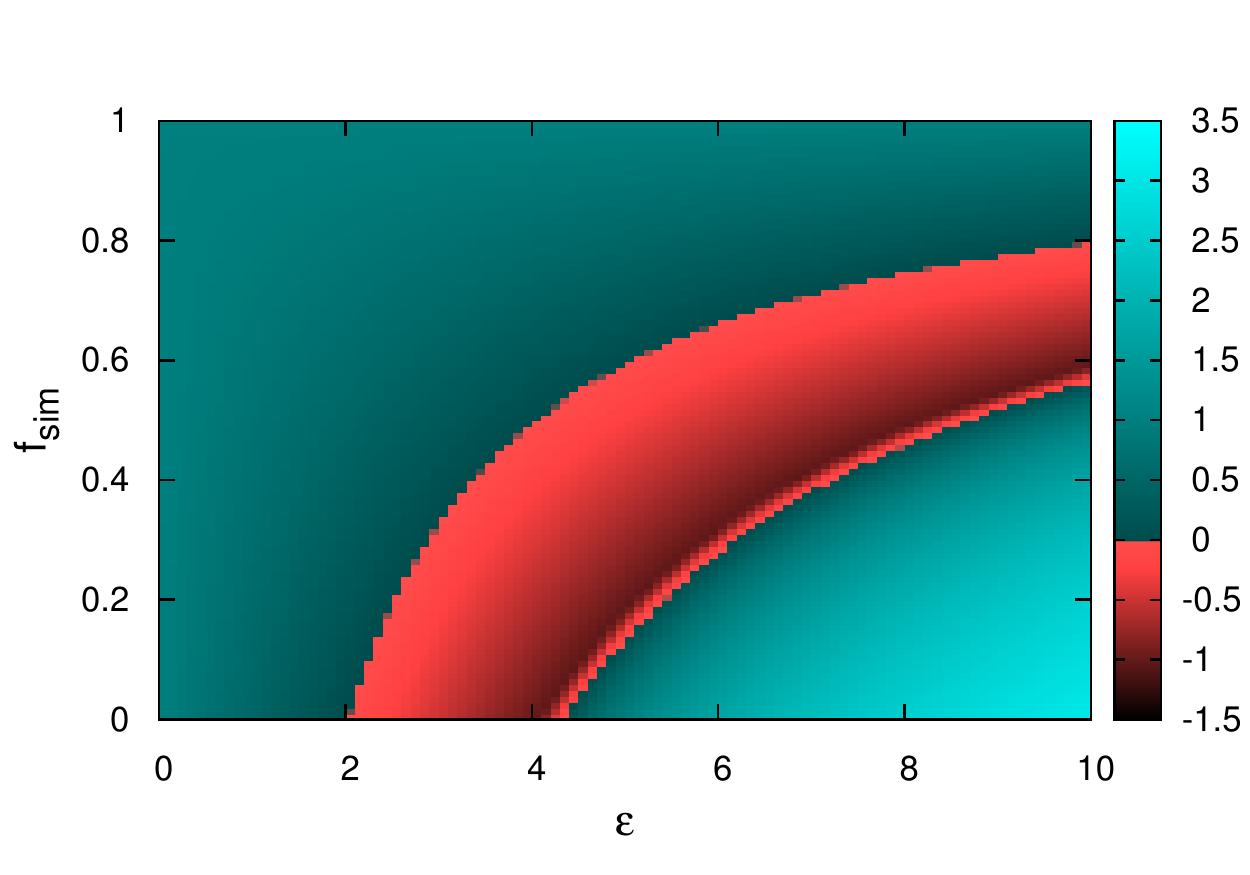}
		\caption{Results from Linear Stability analysis of the effective model of the coupled oscillators given by  Eqn.~\ref{eqn_effective}, in the parameter space of coupling strength $\varepsilon$ and $f_{sim}$. The color code represents the value of the maximum eigen value of the Jacobian corresponding to Eqn.~\ref{eqn_effective}, $\lambda_{max}$. The region in pink, where $\lambda_{max} < 0$, represents stable fixed points. The region in green, where $\lambda_{max} > 0$, represents unstable fixed points. Notice the marked similarity with Figs.~\ref{fig:bs_periodic_fixed_t_switch}a and \ref{fig:bs_probabilistic_fixed_t_switch}a.}
		\label{fig:linear_stability_effective}
	\end{figure}

        Increasing the coupling strength, makes the fixed point at
        zero in Eqn.~\ref{eqn_effective} unstable, and leads to
        creation of new fixed
        points. Fig.~\ref{fig:linear_stability_effective} shows
        results from the linear stability analysis of
        Eqn.~\ref{eqn_effective}, in the neighbourhood of the fixed
        point at zero.  The region in pink represents the stable fixed
        point, where the maximum eigen value of the Jacobian
        corresponding to Eqn.~\ref{eqn_effective}, $\lambda_{max}$, is
        negative. The region in green represents unstable fixed
        points, as $\lambda_{max}>0$ there. Notice the marked
        similarity of these results with
        Fig.~\ref{fig:bs_periodic_fixed_t_switch}a (or equivalently
        Fig.~\ref{fig:bs_probabilistic_fixed_t_switch}a), namely the
        fixed point region is completely well-described by the
        analysis of Eqn.~\ref{eqn_effective} when the frequency of
        switching is high ($\sim100$ Hz). Significantly then, the
        results from the global estimates of the basin of stability of
        the fixed point, for rapid periodic and probabilistic
        switching of coupling form, are {\em recovered accurately
          through the linear stability analysis of a set of effective
          dynamical equations.}

\section{Conclusions}

While the variation in links, namely the connectivity matrix, as a
function of time has been investigated in recent times, the dynamical
consequences of time-varying coupling forms is still not
understood. In this work we have explored this new direction in
time-varying interactions, namely we have studied the effect of
switched coupling forms on the emergent behaviour. The test-bed of our
enquiry is a generic system of coupled Stuart-Landau oscillators,
where the form of the coupling between the oscillators switches
between the similar and dissimilar (or conjugate) variables.  We
consider two types of switching, one where the coupling function
changes periodically and one where it changes probabilistically.  When
the oscillators change their coupling form periodically, they are
coupled via similar variables for fraction $f_{sim}$ of the cycle,
followed by coupling to dissimilar variables for the remaining
part. In the case of probabilistic switching, the probability for the
oscillators to be coupled via similar variables is $p_{sim}$, and the
probability of coupling mediated via dissimilar variables is
$(1-p_{sim})$.

We find that time-varying coupling forms suppress oscillations in a
window of coupling strengths, with the window increasing with the
frequency of switching. That is, {\em more rapid changes in coupling
  form leads to large windows of oscillation suppression}, with the
window of amplitude death saturating after a high enough switching
frequency. Interestingly, for low $f_{sim}$ ($p_{sim}$), the {\em
  oscillations are revived} again beyond this window. That is, too low
or too high coupling strengths yield oscillations, while coupling
strengths in-between suppress oscillations.

Also interestingly, the width of the coupling strength window
supporting oscillation suppression is non-monotonic with respect to
$f_{sim} (p_{sim})$, and has a maximum at $f_{sim} (p_{sim}) \sim
0.58$.  Namely, the largest window of fixed point dynamics arises
where there is balance in the probability of occurrence of the
coupling forms.

Focusing on the dependence of the window of oscillation suppression,
at fixed coupling strengths and varying predominance of coupling
forms, we observe the following: for rapidly switched coupling forms,
at large coupling strengths, the oscillation suppression occurs at an
{\em intermediate} value of $f_{sim}$ ($p_{sim}$). Namely, as the
dominance of similar-variable coupling increases the oscillations are
first suppressed, and then after a point the oscillations are revived
again. The fixed point window shifts towards a higher probability of
similar-variable coupling, as coupling strength increases. This is
counter-intuitive, as purely similar-variable coupling yields
oscillatory behaviour, while dissimilar-variable coupling supports
oscillation suppression.

Lastly, we have suggested an effective dynamics that successfully
yields the observed behaviour for rapidly switched coupling forms,
including an accurate estimate of the fixed point window through
stability analysis. Thus our results will potentially enhance the
broad understanding of coupled systems with time-varying connections.


\end{document}